# Using Code Snippets to Teach Programming Languages

Akingbade, Joshua[a]    Yang, Jianhua[b]    Seyedebrahimi, Mir[b]

Department of Computer Science, University of Warwick, UK[a]
WMG, University of Warwick, UK[b]

Corresponding Author's Email: Jianhua.Yang@warwick.ac.uk

## SUMMARY

Coding is a fundamental skill required in the engineering discipline, and much work exists exploring better ways of teaching coding in the higher education context. In particular, Code Snippets (CSs) are approved to be an effective way of introducing programming language units to students. CSs are portions of source code of varying size and content. They can be used in a myriad of ways, one of which is to teach the code they contain as well as its function. To further explore the use of CSs, a pedagogical summer internship project was set up at the Warwick Manufacturing Group (WMG). The scope of the considerations for the study derives from an educational standpoint. Within the evaluations made, the focus was primarily given to pieces of information which proved to provide evidence pertaining to the methodology involved in either teaching or developing teaching materials. By taking the results produced into account from a pedagogical perspective, it was found that several qualities of popular code snippet tutorials which benefit or hinder the learning process, including code length, interactivity, further support, and quality of explanation. These qualities are then combined and used to present a plan for the design of an effective learning resource which makes use of code snippets.

## INTRODUCTION

Code Snippets (CSs) are portions of source code of varying size and content. They can be used in a myriad of ways, one of which is to teach the code they contain as well as its function. Many teaching resources make use of this application to teach coding concepts. An example of such resources is the Learn Python 2 course offered by (Codecademy, 2021).

In this paper, we will assess resources used to teach coding via the use of CSs. The results of these assessments will then be used to put forward guidelines for the development of an efficient and effective teaching resource.

The scope of the considerations for this paper derives from an educational standpoint. Within the evaluations made, focus was primarily given to pieces of information which proved to provide evidence pertaining to the methodology involved in either teaching or developing teaching materials. By taking the results produced into account from a pedagogical perspective, we are able to filter out data which is irrelevant to the central topic e.g., specifications about the languages themselves.

## LITERATURE REVIEW

The use of Computer Simulations for teaching coding closely aligns with the 'Based on a model' methodology proposed by Szlávi et al (2003). This connection is evident when examining the teaching processes involved. The approach involves introducing a function to users, allowing them to understand the concept through method study. Students modify the function, generating new programs, assessing outputs, and adjusting expectations and inputs accordingly (Szlávi and Zsakó, 2003). While slight variations exist in different resources, the underlying procedure remains similar.

Simulations are increasingly recognized as a valuable tool for conducting experiments using computers. This approach aligns with Kolb's Experiential Learning Cycle (Kolb, 2014). Research, particularly that conducted by Govender and Govender (2023), has demonstrated that experiential learning methodologies can be effectively integrated into the context of coding education. This integration facilitates the learning process, fostering the development of knowledge. Their proposed method involves initiating the learning cycle with a concrete experience. This is followed by reflective observation, leading to abstract conceptualization. The cycle culminates in active experimentation, specifically in the context of learning to program a robot, thereby reinforcing the underlying concepts of computer programming.

Teaching coding via simulations, like teaching coding in general, relies on various essential factors for effectiveness. The success of a teaching method is influenced by how it's structured and applied in different contexts. For instance, using Lego Mindstorms to teach programming is effective for audiences where motivation and engagement are key factors (Powers et al., 2006). With numerous methodologies available, tailored efficiency is crucial.

Although factors differ in significance across contexts, many are common across instances. Therefore, tutorials and simulation-based tools share distinguishable characteristics. Authors often adopt similar methods, as seen in an interview study by Head et al (2020), where experienced authors constructing educational documents exhibited common approaches. Regular themes in tutorial writing are indicative of common occurrences in diverse audiences.

# DATA SOURCES

To help distinguish the beneficial and non-beneficial characteristics of teaching resources which use CSs, an analysis was made of popular resources (Table 1). The findings of this analysis were also compared with information gathered from published text on related concepts in order to confirm or disprove the existence of any patterns. The analysis was kept representative by considering resources of different natures, i.e., from written textbooks to videos and web-courses. Their various features and methodologies were documented, alongside the similarities and differences between those of note.

*Table 1. A Classification of the resources assessed*

| Classification | Resource |
|---|---|
| Websites / Web courses | Codecademy, FreeCode Camp, Udacity |
| Books / Textbooks | Think Python, Learn Python 3 the Hard Way, Python Crash Course, C++ Primer, Programming: Principles and Practice Using C++ |
| Videos / Tutorials | FreeCode Camp, Python for beginners, C++ Tutorial for Beginners |

# DISCUSSION

**Length of Snippets Used**

With CSs being the most integral as well as potentially confusing part of any computer science tutorial, intuitively, there exists an emphasis on keeping them as understandable as possible. Most authors of the tutorials offered that they usually incorporate small amounts of self-contained units of code (Head et al., 2020). They also added elements of styling to important portions of code to highlight it.

The results of these interviews correspond with the findings of the assessment of popular tutorials, with over 83% of the assessed tutorials making use of notably short snippets of code such as those illustrated in Figure 1. Additionally, this characteristic of short snippets is also an element of the "tiered language tools" method proposed by Powers et al (2006) as a method of teaching introductory programming.

*Figure 1 Code snippets adapted from Codecademy (2021) tutorials. (a) C++ (b) Python*

```cpp
for (int i = 0; i < 20; i++)
{
    std::cout << "I will not throw paper airplanes in class.\n";
}
```

```python
class Square(object):
  def __init__(self):
    self.sides = 4

my_shape = Square()
```

*(a)*        *(b)*

The snippets are kept short to ensure minimal syntax and complexity which help students to not only focus on understanding the language being taught but also successfully follow the tutorial (Head et al., 2020, Powers et al., 2006).

**Interactivity**

As essential as the content of the tutorial is, without a means of grasping and maintaining a hold on the student's attention, there exists a large risk of pupils not following the tutorial completely. Tutorial authors seemed to understand this, answering that they designed and revised tutorials to ensure they could hold a reader's attention (Head et al., 2020). These answers were also supported by "visual programming tools" and "specialised realization" teaching methods in which emphasis is laid on giving a hands-on interactive experience during teaching (Powers et al., 2006). This concept is also further supported by other teaching methods which use similar methods to increase motivation of students when learning.

Most of the assessed tutorials offered some form of interactivity with the content being taught much like those displayed in Figure 2. This ranged from the exercises put forward by textbook materials (Lippman et al., 2013, Downey, 2008, Shaw, 2017, Stroustrup, 2014), to the interactive terminal tools offered by some of the web-courses in which immediate visual feedback is given, much like the previously highlighted specialized realization method (Powers et al., 2006).

Figure 2. Interactive mediums used by popular tutorials (a) Codecademy (2021). (b) FreeCodeCamp (2021)

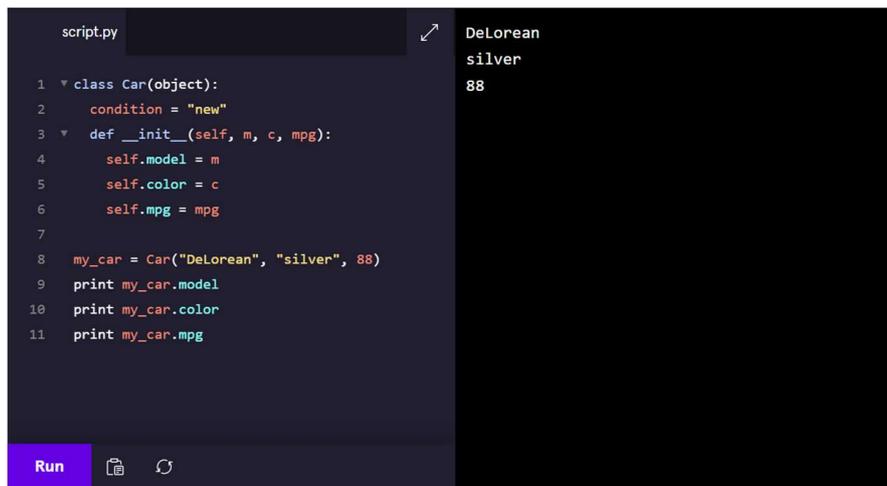

(a)

(b)

## Larger Project

Much like the previously discussed features, the interviewed authors also highlighted the use of an overarching project, the source code for which was broken up to form the constituting CSs (Head et al, 2020). The use of an end-project provides the user with an application of the concept which is being learnt. Evidence supporting this ideology is showcased by the online criticisms of computer science tutorial users which voice a lack of knowledge of the application of the taught content (KokotsakiMenzies and Wiggins, 2016).

A large percentage of the reviewed tutorials do not incorporate a means with which the user may apply knowledge of the coding concept. The small number of those which did use a final project made sure the constituting CSs came together to form the said final project. This allowed for the smooth development of the project throughout the tutorial by the pupils.

**Further support**

Though the code snippets alongside their accompanying explanations remain the most important aspects of a tutorial, the additional support offered by the teaching resource is also a key constituent. Whereas the CSs, explanations and the like all stem to provide knowledge to the student, the additional support provided is supposed to cater to the student's wellbeing during the teaching process. This should include, but not be limited to, both assisting with the teaching resource as well as offering career advice and even a sense of community.

Most of the reviewed materials and resources either lacked a definitive support system (most web courses) or did not have one entirely (all textbooks). It was found that most of the web tutorials offered help in the form of problem hints coupled with access to an online forum, though its usefulness was heavily limited due to the inactivity of its members. From polling user reviews, it was also found that complaints were made over the absence of aid for technical matters in web courses (Trustpilot, 2021).

**Quality of Explanations**

The prose explanations used in the tutorials are another crucial element as they give definition to the CSs which they accompany. Tutorial authors gave insight into their routine, offering that they sought to keep text brief and clear (Head et al., 2020). Some interviewees also explained that bodies of text were broken up through use of photos, CSs as well as other mediums. Results from the reviewing of popular tutorials reinforces these claims as many of them integrate videos and even memes to prevent tutorials from being dry and unengaging whilst conveying motivation unto the user.

**COURSE DESIGN RECOMMENDATIONS**

For the purpose of designing a computer science tutorial, we shall take into account the results of reviewing well known tutorials. This is done with the intent of incorporating positive noted features while avoiding the negative, as shown in Figure 3

- High Quality Code Snippets: To begin with, the CSs used within the tutorial must be of a high quality. This entails not only ensuring that the snippets are short in length, but also that they are not overly complex. Snippets should be self-contained units of code, carrying out one specific function for which they are designed.
- High Quality Prose: As well as the CSs themselves, the prose which accompanies them also must be of exceptional quality. The explaining text should be brief and clear, explaining the contents of the CSs as well as their function in a manner which does not come across as dry and unenergetic. The inclusion of humour also can also assist in making sure the explanation is engaging and motivational for the user.

- Highly Interactive: Additionally, some form of interactivity should be implemented in the resource. Any number of options may be used individually or better yet combined in the tutorial. Ranging from questions which challenge the user's understanding of the concept to interactive terminals and even physical media which would allow for the pupil to receive immediate visual feedback.
- Use of a Final Project: Moreover, a final project should be used to allow for the user to gain knowledge of the application of the taught coding concept. This larger project should consist of the CSs used within the tutorial in order to ensure its development through the completion of the course.
- Additional Support: In addition, the resource should offer support throughout the learning process as this contributes towards the user's well-being. As the resource is essentially assuming the role of a teacher in that it conveys information on the set topic, it should also fulfil the duty of providing the user with technical support, as well as career advice and a forum to assist with any other issues which may crop up.

*Figure 3. Steps in course design (graph generated using templates on Lucid (2021)*

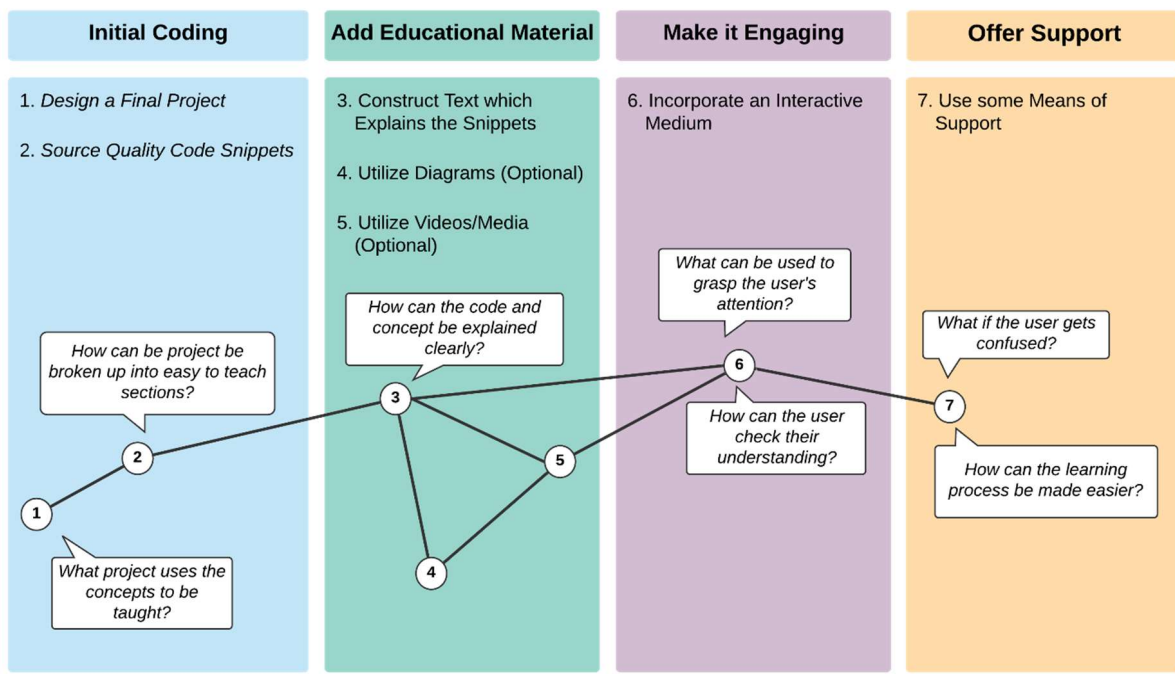

# CONCLUSIONS

In conclusion, we believe that the factors surrounding the implementation of a computer science tutorial play an important role in its impact and effectiveness on prospective users. The results of assessing various popular learning resources yielded there exist several components, of which some are beneficial, and others are less so. Using these results, we have outlined means through which an effective teaching asset can be developed. Overall, these guidelines aim to produce a resource which is not only impactful in the teaching of simple coding concepts, but also for more complex topics.